\documentclass[twocolumn]{revtex4}

\usepackage{graphicx}
\usepackage{dcolumn}
\usepackage{bm}

\input epsf

\begin{document}

\title{\Large Brans-Dicke Theory and Thermodynamical Laws on Apparent and Event Horizons}

\author{\bf Samarpita Bhattacharya\footnote{samarpita$_{-}$sarbajna@yahoo.co.in} and Ujjal Debnath\footnote{ujjaldebnath@yahoo.com ,
ujjal@iucaa.ernet.in}}

\affiliation{Department of Mathematics, Bengal Engineering and
Science University, Shibpur, Howrah-711 103, India.}

\date{\today}

\begin{abstract}
In this work, we have described the Brans-Dicke theory of gravity
and given a particular solution by choosing a power law form of
scalar field $\phi$ and constant $\omega$. If we assume first law
and entropy formula on apparent horizon then we recover Friedmann
equations. Next, assuming first law of thermodynamics, the
validity conditions of GSL on event horizon are presented. Also
without use first law, if we impose the entropy relation on the
horizon, then we also obtain the condition of validity of GSL on
event horizon. The validity of GSL completely depends on the model
of BD scalar field solutions. We have justified that on the
apparent horizon the two process are equivalent, but on the event
horizon they are not equivalent. If first law is valid on the
event horizon then GSL may be satisfied in BD solution, but if
first law is not satisfied then GSL is not satisfied in BD
solution. So first law always favours GSL on event horizon. In our
effective approach, the first law and GSL is always satisfied in
apparent horizon, which do not depend on BD theory of gravity.
\end{abstract}


\maketitle

\section{\normalsize\bf{Introduction}}

The luminosity-redshift relation observed for type-Ia supernovae
[1, 2] strongly suggests that, in the present phase, the universe
is undergoing an accelerated expansion. This is supported also by
recent measurements of CMBR and the power spectrum of mass
perturbations. There are several proposals regarding this, it
cosmological constant and quintessence like dark energy [3 - 5]
being some of the competent candidates. Basically quintessence is
a dynamical slowly evolving spatially inhomogeneous component of
energy density with negative pressure. The vector and tensor
fields describing the fundamental forces, there may exist scalar
field. The energy density associated with a scalar field $\phi$
slowly moving down its potential $V$ can represent a simple
example of quintessence. Another simplest alternative which
includes the scalar field in addition to the tensor field in
general relativity is Brans-Dicke (BD) theory. Brans-Dicke theory
has been proved to be very effective regarding the recent study of
cosmic acceleration [6]. BD theory is explained by a scalar
function $\phi$ and a constant coupling constant $\omega$, often
known as the BD parameter. This can be obtained from general
theory of relativity (GR) by letting $\omega \rightarrow \infty$
and $\phi=$ constant [7]. This theory has very effectively solved
the problems of inflation and the early and the late time
behaviour of the Universe. BD scalar-tensor theory can potentially
solve the quintessence problem [6]. The generalized BD theory [8]
is an extension of the original BD theory with coupling function
$\omega$ is a function of the scalar field $\phi$. Bertolami and
Martins [9] have used this theory to present an accelerated
Universe for spatially flat model. All these theories conclude
that $\omega$ should have a low negative value in order to solve
the cosmic acceleration problem. This contradicts the solar system
experimental bound $\omega\geq 500$. However they have obtained
the solution for accelerated expansion with a potential
${\phi}^{2}$ and large $|\omega|$, although they have not
considered the positive energy conditions for the matter and
scalar field. In context of accelerating expansion of the
universe, there are several works on BD theory [10, 11] in both
theoretical and observational point of views.\\

Motivated by the black hole physics, it was realized that there is
a profound connection between gravity and thermodynamics. In
Einstein gravity, the evidence of this connection was first
discovered in [12] by deriving the Einstein equation from the
proportionality of entropy and horizon area together with the
first law of thermodynamics in the Rindler spacetime. The horizon
area (geometric quantity) of black hole is associated with its
entropy (thermodynamical quantity), the surface gravity (geometric
quantity) is related with its temperature (thermodynamical
quantity) in black hole thermodynamics [13]. In 1995, Jacobson
[12] was indeed able to derive Einstein equations by applying the
first law of thermodynamics $\delta Q = TdS$ together with
proportionality of entropy to the horizon area of the black hole.
He assumed that this relation holds for all Rindler causal
horizons through each space time point with $\delta Q$ and $T$
interpreted as the energy flux and temperature seen by an
accelerated observer just inside the horizon. Then Padmanabhan
[14] was able to formulate the first law of thermodynamics on the
horizon, starting from Einstein equations for a general static
spherically symmetric space time.\\

Frolov and Kofman in [15] employed the approach proposed by
Jacobson [12] to a quasi-de Sitter geometry of inflationary
universe, where they calculated the energy flux of a background
slow-roll scalar field (inflaton) through the quasi-de Sitter
apparent horizon and used the first law of thermodynamics $-dE =
TdS$, where $dE$ is the amount of the energy flow through the
apparent horizon. Although the topology of the local Rindler
horizon in Ref. [12] is quite different from that of the quasi-de
Sitter apparent horizon considered in Ref. [15], it was found that
this thermodynamic relation reproduces one of the Friedmann
equations with the slow-roll scalar field. It is assumed in their
derivation that $T=\frac{H}{2\pi}$ and $S=\frac{\pi}{GH^{2}}$
where $H$ is a slowly varying Hubble parameter. Also the identity
between Einstein equations and thermodynamical laws has been
applied in the cosmological context considering universe as a
thermodynamical system bounded by the apparent horizon (RA). Using
the Hawking temperature $T_{A}=\frac{1}{2\pi R_{A}}$ and
Bekenstein entropy $S_{A}=\frac{\pi R_{A}^{2}}{G}$ ($R_{A}$ is the
radius of apparent horizon) at the apparent horizon, the first law
of thermodynamics (on the apparent horizon) is shown to be
equivalent to Friedmann equations [16] and the generalized second
law of thermodynamics (GSLT) is obeyed at the horizon.  The
thermodynamics in de Sitter space–time was first investigated by
Gibbons and Hawking in [17]. In a spatially flat de Sitter
space–time, the event horizon and the apparent horizon of the
Universe coincide and there is only one cosmological horizon. In
the usual standard big bang model a cosmological event horizon
does not exist. But for the accelerating universe dominated by
dark energy, the cosmological event horizon separates from that of
the apparent horizon. When the apparent horizon and the event
horizon of the Universe are different, it was found that the first
law and generalized second law (GSL) of thermodynamics hold on the
apparent horizon, while they break down if one considers the event
horizon [18]. On the basis of the well known correspondence
between the Friedmann equation and the first law of thermodynamics
of the apparent horizon, Gong et al [19] argued that the apparent
horizon is the physical horizon in dealing with thermodynamics problems.\\

Recently, it is of great interest to study the generalized second
law (GSL) of thermodynamics in the generalized gravity theories.
There have been a lot of interest on investigating the GSL in
gravity [20,21], but all of them concentrate on the Einstein
gravity. The modified theory of gravity was argued to be a
possible candidate to explain the accelerated expansion of our
universe, thus it is interesting to examine the GSL in the
extended gravity theories. Even for the Einstein gravity, it was
found that GSL breaks down in phantom-dominated universe in the
presence of Schwarzschild black hole [22]. Entropy of the horizon
from the first law of thermodynamics constructed in [20]. The
total entropy evolution with time including the horizon entropy,
the non-equilibrium entropy production, and the entropy of all
matter, field and energy components have been discussed by Wu et
al [23]. They have shown a universal condition to protect the GSL
in generalized gravity theories and its validity in the Einstein
gravity (even in the presence of Schwarzschild black hole) and
higher order gravity. Also there are several works on
thermodynamics due to Gauss-Bonnet gravity [24], Horava-Lifshitz
gravity [25], Lovelock gravity [16, 23, 26], braneworld gravity
[23, 26, 27], $f(R)$ gravity [23, 28, 29] and scalar-tensor
gravity [23, 28]. \\

In this work, we briefly describe the Brans-Dicke theory of
gravity and give a particular solution by choosing a power law
form of scalar field $\phi$ and constant $\omega$ in section II.
Next, assuming first law of thermodynamics, the validity
conditions of GSL on event horizon are presented in section III.
Also without use first law if we impose the entropy relation on
the horizon, then we also obtain the condition of validity of GSL
on event horizon in section IV. There are two ways to get validity
conditions of GSL on apparent and event horizons: (i) use first
law and find entropy relation on the horizons (ii) use only
horizon entropy on the horizons. Finally some concluding remarks
have been presented in last section.\\

\section{\normalsize\bf{Brans-Dicke Theory}}

The self-interacting Brans-Dicke (BD) theory is described by the
Jordan-Brans-Dicke (JBD) action [30] : (choosing $c=1$)

\begin{equation}
S=\int\frac{d^{4} x \sqrt{-g}}{16\pi}\left[\phi R-
\frac{\omega(\phi)}{\phi} {\phi}^{,\alpha}
{\phi,}_{\alpha}-V(\phi)+ 16\pi{\cal L}_{m}\right]
\end{equation}

where $V(\phi)$ is the self-interacting potential for the BD
scalar field $\phi$ and $\omega(\phi)$ is modified version of the
BD coupling parameter which is a function of $\phi$. In this
theory $\frac{1}{\phi}$ plays the role of the gravitational
constant $G$. This action also matches with the low energy string
theory action [31] for $\omega=-1$. The matter content of the
Universe is composed of matter fluid, so the energy-momentum
tensor is given by

\begin{equation}
T_{\mu \nu}^{m}=(\rho+p)u_{\mu} u_{\nu}+p~g_{\mu \nu}
\end{equation}

where $u^{\mu}$ is the four velocity vector of the matter fluid
satisfying $u_{\mu}u^{\nu}=-1$ and $\rho,~p$ are respectively
energy density and isotropic pressure.\\

From the Lagrangian density $(1)$ we obtain the field equations
[11]

\begin{eqnarray*}
G_{\mu \nu}=\frac{8\pi}{\phi}T_{\mu
\nu}^{m}+\frac{\omega(\phi)}{{\phi}^{2}}\left[\phi  _{ , \mu} \phi
_{, \nu} - \frac{1}{2}g_{\mu \nu} \phi _{, \alpha} \phi ^{ ,
\alpha} \right]~~~~~~~~~~~~~~~~~~~
\end{eqnarray*}
\begin{equation}
 +\frac{1}{\phi}\left[\phi  _{, \mu ; \nu} -g_{\mu
\nu}~ ^{\fbox{}}~ \phi \right]-\frac{V(\phi)}{2 \phi} g_{\mu \nu}
\end{equation}

and

\begin{eqnarray*}
^{\fbox{}}~\phi=\frac{8\pi
T}{3+2\omega(\phi)}-\frac{1}{3+2\omega(\phi)}\left[2V(\phi)-\phi
 \frac{dV(\phi)}{d\phi}\right]
\end{eqnarray*}
\begin{equation}
~~~~~~~~~~~~~~~-\frac{\frac{d\omega(\phi)}{d\phi}}{3+2\omega(\phi)}{\phi,}_{\mu}
 {\phi}^{,\mu}
\end{equation}

where $T=T_{\mu \nu}^{m}g^{\mu \nu}$. Equation (3) can also be written as

\begin{equation}
G_{\mu\nu}=8\pi
\tilde{T}_{\mu\nu}=\frac{8\pi}{\phi}(T_{\mu\nu}^{m}+\frac{1}{8\pi}T_{\mu\nu}^{\phi})
\end{equation}

where $\tilde{T}_{\mu\nu}$ can be treated as effective energy
momentum tensor. The line element for Friedman-Robertson-Walker
space-time is given by

\begin{equation}
ds^{2}=-dt^{2}+a^{2}(t)\left[\frac{dr^{2}}{1-kr^{2}}+r^{2}(d\theta^{2}+sin^{2}\theta
d\phi^{2})\right]
\end{equation}
\\
where, $a(t)$ is the scale factor and $k~(=0, -1,+1)$ is the
curvature index describe the flat, open and closed model of the universe.\\

The Einstein field equations and the wave equation for the BD
scalar field $\phi$ are in the following [11]

\begin{equation}
H^{2}+\frac{k}{a^{2}}=\frac{8\pi\rho}{3\phi}-H\frac{\dot{\phi}}{\phi}+\frac{\omega(\phi)}{6}\frac{\dot{\phi}^{2}}{\phi
^{2}}+\frac{V(\phi)}{6\phi}
\end{equation}

\begin{equation}
 2\dot{H}+3H^{2}+\frac{k}{a^{2}}=-\frac{8\pi p}{\phi}-\frac{\omega(\phi)}{2}\frac{\dot{\phi}^{2}}{\phi
^{2}}-2H\frac{\dot{\phi}}{\phi}-\frac{\ddot{\phi}}{\phi}+\frac{V(\phi)}{2\phi}
\end{equation}

and

\begin{eqnarray*}
\ddot{\phi}+3H
\dot{\phi}=\frac{8\pi(\rho-3p)}{3+2\omega(\phi)}+\frac{1}{3+2\omega(\phi)}\left[2V(\phi)-\phi
\frac{dV(\phi)}{d\phi}\right]
\end{eqnarray*}
\begin{equation}
~~~~~~~~~~~~~~~~~~~~~~~~~~~~~~~~~~~~~-\frac{\frac{d\omega(\phi)}{d\phi}}{3+2\omega(\phi)}
\end{equation}

where $H=\frac{\dot{a}}{a}$ is the Hubble parameter. Now let us
assume the matter is conserved in BD theory. So the matter
conservation equation is given by

\begin{equation}
\dot{\rho}+3H(\rho+p)=0
\end{equation}

Equations (7) and (8) can be written as

\begin{equation}
H^{2}+\frac{k}{a^{2}}=\frac{8\pi}{3}\rho_{eff}
\end{equation}

and

\begin{equation}
 2\dot{H}+3H^{2}+\frac{k}{a^{2}}=-8\pi p_{eff}
\end{equation}

where $\rho_{eff}$ and $p_{eff}$ are effective fluid density and
pressure for combination of matter and the contribution of BD
field respectively defined  by

\begin{equation}
\rho_{eff}=\frac{\rho}{3\phi}+\frac{3}{8\pi}\left(
-H\frac{\dot{\phi}}{\phi}+\frac{\omega(\phi)}{6}\frac{\dot{\phi}^{2}}{\phi
^{2}}+\frac{V(\phi)}{6\phi}\right)
\end{equation}

\begin{equation}
p_{eff}=\frac{p}{\phi}+\frac{1}{8\pi}\left(
\frac{\omega(\phi)}{2}\frac{\dot{\phi}^{2}}{\phi
^{2}}+2H\frac{\dot{\phi}}{\phi}+\frac{\ddot{\phi}}{\phi}-\frac{V(\phi)}{2\phi}\right)
\end{equation}

From (11) and (12), we see that the field equations are same as
the usual Friedmann equations in Einstein gravity. If we use the
wave equation (9) and the matter conservation equation (10), we
obviously obtain the conservation equation of effective fluid as

\begin{equation}
\dot{\rho}_{eff}+3H(\rho_{eff}+p_{eff})=0
\end{equation}

Here we consider the Universe to be filled with barotropic fluid
with EOS

\begin{equation}
p=w \rho~~~(-1<w<1)
\end{equation}

The conservation equation $(10)$ yields the solution for $\rho$
as,
\begin{equation}
\rho=\rho_{0} a^{-3(1+w)}
 \end{equation}

where $\rho_{0}(>0)$ is an integration constant.\\

Let us choose $\omega(\phi)=\omega=$ constant. In this case we
consider only one power law form of $\phi$ as

\begin{equation}
\phi=\phi_{0} a^{\alpha}
\end{equation}

Using equation $(16)$ and (17) and solving equations $(7)$ and
$(8)$ we get

\begin{equation}
H^{2}=k A a^{-2}+B
a^{-\frac{2\alpha((1+\omega)\alpha-1)}{2+\alpha}}+Ca^{-\alpha-3(1+w)}
\end{equation}

\begin{equation}
\dot{H}=k A_{1} a^{-2}+B_{1}
a^{-\frac{2\alpha((1+\omega)\alpha-1}{2+\alpha}}+C_{1}a^{-\alpha-3(1+w)}
\end{equation}

and

\begin{equation}
V=k A_{2} a^{\alpha-2}+B_{2}
a^{\alpha-\frac{2\alpha((1+\omega)\alpha-1}{2+\alpha}}+C_{2}a^{-3(1+w)}
\end{equation}

where $A=\frac{2}{\alpha((1+\omega)\alpha-1)-(2+\alpha)}$,
$C=-\frac{16\pi(1+w)\rho_{0}}{\phi_{0}[(2+\alpha)(-\alpha-3(1+w))+2\alpha((1+\omega)\alpha-1
)]}$, $A_{1}=\frac{2+\alpha(1-\alpha(1+\omega))A}{2+\alpha}$,
$B_{1}=\frac{\alpha(1-\alpha(1+\omega))B}{2+\alpha}$,
$C_{1}=\frac{\alpha(1-\alpha(1+\omega))\phi_{0}C-8\pi(1+w)\rho_{0}
}{(2+\alpha)\phi_{0}}$, $A_{2}=6\phi_{0}+AB_{2}/B$,
$B_{2}=6\pi_{0}(1+\alpha-\omega\alpha^{2}/6)B $, $C_{2}=B_{2}C/B-16\pi\rho_{0}$
and $B$ is an arbitrary constant.\\

\section{\normalsize\bf{Study of Thermodynamics on Apparent and Event Horizons}}

In this section, we assume that the first law is valid [32] on
apparent/event horizons and after that we examine the validity of
GSL on apparent/event horizons.\\

From the first law of thermodynamics, we get the relation [16, 33]

\begin{eqnarray*}
T_{X}dS_{X}=-dE_{X}=4\pi
R_{X}^{3}H\tilde{T}_{\mu\nu}k^{\mu}k^{\nu}dt
\end{eqnarray*}
\begin{equation}
~~~~~~~~~~~~~~~~~~~=4\pi R_{X}^{3}H(\rho_{eff}+p_{eff})dt
\end{equation}

where suffix $X$ denotes the apparent horizon ($X=A$) and event
horizon ($X=E$). Also $T_{X}$ and $R_{X}$ are the temperature and
radius of apparent/event horizons. The radii of apparent and event
horizons are defined by

\begin{equation}
R_{A}=\frac{1}{\sqrt{H^{2}+\frac{k}{a^{2}}}}
\end{equation}
and

\begin{equation}
R_{E}=a\int_{a}^{\infty}\frac{da}{a^{2}H}
\end{equation}

which immediately give,

\begin{equation}
\dot{R}_{A}=-HR_{A}^{3}\left(\dot{H}-\frac{k}{a^{2}} \right)
\end{equation}
and
\begin{equation}
\dot{R}_{E}=HR_{E}-1
\end{equation}

Now from equation (22) we get the rate of change of entropy on the
apparent/event horizon as

\begin{equation}
\dot{S}_{X}=\frac{4\pi R_{X}^{3}H}{T_{X}}(\rho_{eff}+p_{eff})
\end{equation}

To study GSL of thermodynamics through the universe we deduce the
expression for normal entropy using the Gibb's equation of
thermodynamics [18]

\begin{equation}
T_{X}dS_{IX}=p_{eff}dV+d(E_{IX})
\end{equation}

where, $S_{IX}$ is the internal entropy within the apparent/event
horizon. Here the expression for internal energy can be written as
$E_{IX}=\rho_{eff} V_{X}$, where the volume of the sphere is
$V_{X}=\frac{4}{3}\pi R_{X}^{3}$. So from equation (28), we obtain
the rate of change of internal energy as

\begin{equation}
\dot{S}_{IX}=\frac{4\pi
R_{X}^{2}}{T_{X}}(\rho_{eff}+p_{eff})(\dot{R}_{X}-HR_{X})
\end{equation}

Adding (27) and (29), the rate of change of total entropy is
obtained as

\begin{equation}
\dot{S}_{tX}=\dot{S}_{X}+\dot{S}_{IX}=\frac{4\pi
R_{X}^{2}\dot{R}_{X}}{T_{X}}(\rho_{eff}+p_{eff})
\end{equation}

Using (11), (12), (25) and (30), we see that on the apparent
horizon $\dot{S}_{tA}\ge 0$ always. So on the apparent horizon the
GSL is satisfied. So validity of GSL on apparent horizon do not
depend on BD theory of gravity. But on the event horizon the GSL
will be satisfied if any one of the following conditions hold: (i)
$\dot{R}_{E} \ge 0$ and $\rho_{eff}+p_{eff} \ge 0$ or (ii)
$\dot{R}_{E} \le 0$ and $\rho_{eff}+p_{eff} \le 0$. Here, the
inequalities completely depend on BD scalar field solutions. So
validity of GSL on event horizon completely depends on BD theory
of gravity. From (13) and (14), we see that ($\rho_{eff}+p_{eff}$)
is independent of $V$. So the validity of GSL depends only on
matter and scalar field, but not its potential. Now using BD
solutions (17)-(20), we draw the rate of change of total entropy
on the event horizon i.e., $\dot{S}_{tE}$ against $z$ in figure 1
for open, closed and flat models. From the figure it is to be seen
that when $z$ decreases, $\dot{S}_{tE}$ becomes positive. So we
conclude that BD field supports to GSL of thermodynamics.\\

\begin{figure}
\includegraphics[height=2in]{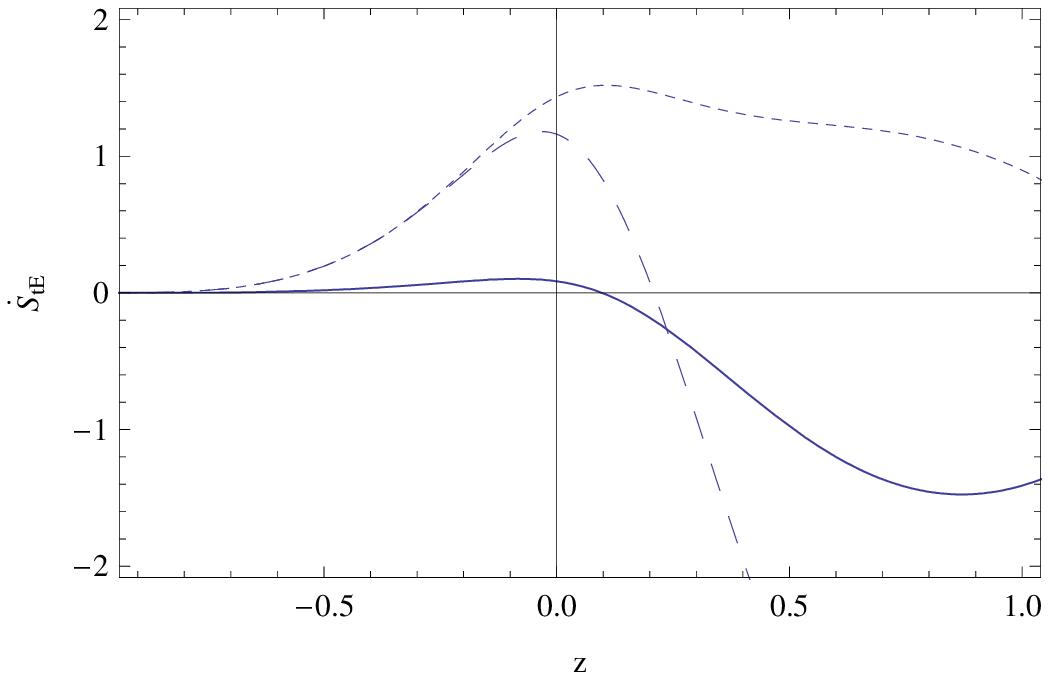}~~~~

\vspace{6mm} Fig. 1 represents the variation of $\dot{S}_{tE}$
(equation (30)) against redshift $z$ for $w=-2/3,\omega=-10$ and
$k=0,\pm 1$. The dashed line, dotted line and filled line
represent for $k=0,~-1$ and $+1$ respectively.

\vspace{6mm}

\end{figure}

\section{\normalsize\bf{Study of Thermodynamics on the Event Horizon without using First Law}}

Here we assume that the first law is valid only on the apparent
horizon. So from equation (27), we have rate of change of entropy
on apparent horizon,

\begin{equation}
\dot{S}_{A}=\frac{4\pi R_{A}^{3}H}{T_{A}}(\rho_{eff}+p_{eff})
\end{equation}

which immediately leads to

\begin{equation}
S_{A}=\int\frac{4\pi R_{A}^{3}H}{T_{A}}(\rho_{eff}+p_{eff})dt
\end{equation}

where radius of apparent horizon is defined in equation (23). Now
we consider the entropy and temperature on apparent horizon in
Einstein's gravity i.e.,

\begin{equation}
S_{A}=\pi R_{A}^{2}
\end{equation}

and

\begin{equation}
T_{A}=\frac{1}{2\pi R_{A}}
\end{equation}

Using (23), (33) and (34), the equation (31) gives us

\begin{equation}
\dot{H}-\frac{k}{a^{2}}=-4\pi(\rho_{eff}+p_{eff})
\end{equation}

Also eliminating $(\rho_{eff}+p_{eff})$ from (15) and (35) and
after integrating we obtain

\begin{equation}
H^{2}+\frac{k}{a^{2}}=\frac{8\pi}{3}\rho_{eff}
\end{equation}

The equations (35) and (36) are the Friedmann equations. So if we
consider the first law of thermodynamics is valid on the apparent
horizon, we recover the Friedmann equations in Einstein's gravity.\\

Another way, if we only consider the entropy defined in (33)
rather than the entropy formula (31) from the first law on the
apparent horizon, then the derivative of the entropy will be

\begin{equation}
\dot{S}_{A}=2\pi R_{A}\dot{R}_{A}
\end{equation}

Also the rate of change of internal entropy on the apparent
horizon is (from (29))

\begin{equation}
\dot{S}_{IA}=\frac{4\pi
R_{A}^{2}}{T_{A}}(\rho_{eff}+p_{eff})(\dot{R}_{A}-HR_{A})
\end{equation}

Adding (37) and (38), and using (11), (12), (25) and (34) and
after manipulation we get the rate of change of total entropy on
the apparent horizon:

\begin{equation}
\dot{S}_{A}+\dot{S}_{IA}=\frac{4\pi
R_{A}^{2}\dot{R}_{A}}{T_{A}}(\rho_{eff}+p_{eff})
\end{equation}

which is same as equation (30) on apparent horizon. So we see that
the rate of change of total entropy on the apparent horizon by
considering first law of thermodynamics is identical with the rate
of change of total entropy by considering the entropy formula on
the apparent horizon. So we may conclude that the entropy formulae
(32) and (33) are equivalent. But on the event horizon, we do not
know that this result may or may not hold. Now using first law,
the rate of change of total entropy formula for the event horizon
has been given in equation (30). In this section we do not
consider the first law of thermodynamics while we only consider
the entropy and temperature on the event horizon in Einstein's
gravity i.e.,

\begin{equation}
S_{E}=\pi R_{E}^{2}
\end{equation}

and

\begin{equation}
T_{E}=\frac{1}{2\pi R_{E}}
\end{equation} \\

Now using (26), (27), (40) and (41), we obtain the rate of change
of total entropy for the event horizon:

\begin{eqnarray*}
\dot{S}_{tE}=\dot{S}_{IE}+\dot{S}_{E}=\frac{4\pi
R_{E}^{2}\dot{R}_{E}}{T_{E}}(\rho_{eff}+p_{eff})
\end{eqnarray*}
\begin{equation}
~~~~~~~~~~~~~~~~~~~~~~~~~+2\pi R_{E}(HR_{E}-1-4\pi R_{E}^{3}H)
\end{equation}

which implies

\begin{eqnarray*}
\dot{S}_{tE}=\dot{S}_{IE}+\dot{S}_{E}=-8\pi^{2}
R_{E}^{3}(\rho_{eff}+p_{eff})
\end{eqnarray*}
\begin{equation}
~~~~~~~~~~~~~~~~~~~~~~~~~~~~~+2\pi R_{E}(HR_{E}-1)
\end{equation}

Comparing equations (30) and (42), we see that there is an extra
term (positive or negative) arises in (42) than (30). So the rate
of change of total entropy on the event horizon by considering
first law of thermodynamics is not identical with the rate of
change of total entropy by considering the entropy formula on the
event horizon. So we may conclude that the entropy formulae (27)
and (40) are not equivalent on the event horizon. Or shortly
speaking, if first law is valid on event horizon then entropy
formula (40) is not valid on the event horizon or, if entropy
formula (40) is valid on the event horizon then first law cannot
be satisfied on the event horizon. If r.h.s of (43) is
non-negative then GSL is satisfied on the event horizon, which
depends on the BD scalar field solutions. From (13) and (14), we
see that ($\rho_{eff}+p_{eff}$) is independent of $V$. So the
validity of GSL depends only on matter and scalar field, but not
its potential. Now using BD solutions (17)-(20), we draw the rate
of change of total entropy on the event horizon i.e.,
$\dot{S}_{tE}$ against $z$ in figure 2 for open, closed and flat
models. From the figure it is to be seen that when $z$ decreases,
$\dot{S}_{tE}$ becomes negative. So we may conclude that GSL of
thermodynamics is not satisfied for the BD solutions. \\

\begin{figure}
\includegraphics[height=2in]{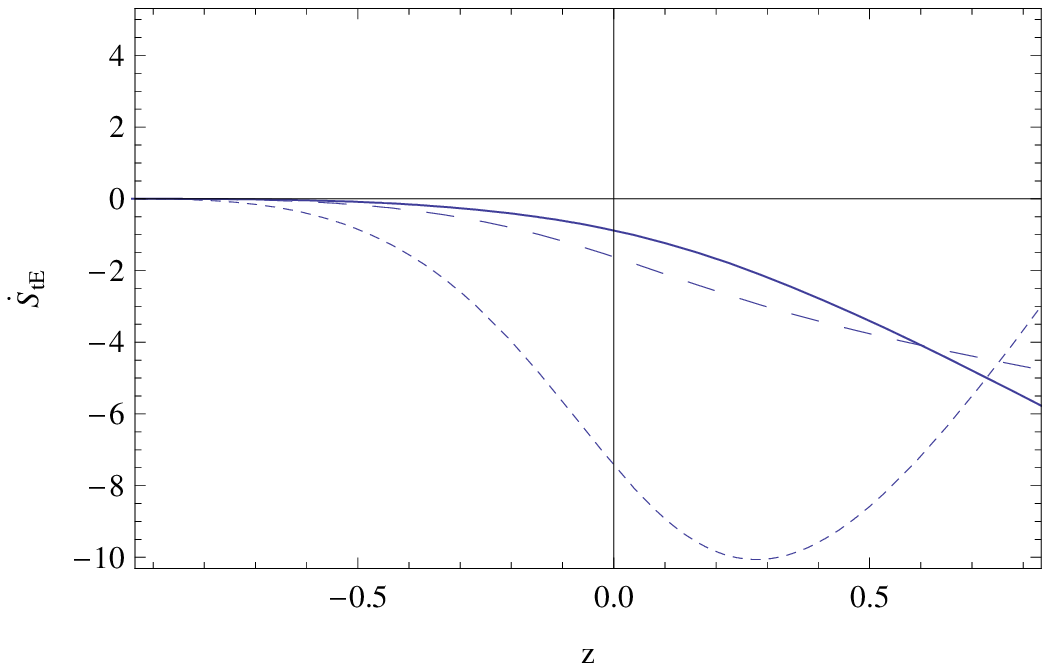}~~~~

\vspace{6mm} Fig. 2 represents the variation of $\dot{S}_{tE}$
(equation (43)) against redshift $z$ for $w=-2/3,\omega=-10$ and
$k=0,\pm 1$. The dashed line, dotted line and filled line
represent for $k=0,~-1$ and $+1$ respectively.

\vspace{6mm}

\end{figure}

\section{\normalsize\bf{Discussions}}

In this work, we have described the modified version of
Brans-Dicke theory of gravity and given a particular solution by
choosing a power law form of scalar field $\phi$ and constant
$\omega$. We have examined the validity of GSL on apparent and
event horizons in BD theory by an effective approach. Next,
assuming first law of thermodynamics, the validity conditions of
GSL on event horizon are presented. If we assume first law and
entropy formula on the apparent horizon then using continuity
equation of matter fluid, we can recoverd two Friedmann equations.
Also without use first law if we impose the entropy relation on
the horizon, then we also obtain the condition of validity of GSL
on th event horizon. There are two ways to get validity conditions
of GSL on apparent and event horizons: (i) use first law and find
entropy relation on the horizons (ii) use horizon entropy on the
horizons. On the apparent horizon the two process are equivalent,
but on the event horizon they are not equivalent. We have also
shown that the GSL on the apparent horizon is always valid in BD
theory also. We have also shown that the GSL do not depend on the
scalar field potential in BD theory. From fig.1 and fig.2, we see
that there are opposite behaviours of $\dot{S}_{tE}$, i.e. if
first law is valid on event horizon then GSL is successfully
satisfied for BD solution, but if first law is not valid then the
GSL is no longer satisfied for BD solution. So we may conclude
that first law always favours GSL of thermodynamics in BD theory.
In our effective approach, the first law and GSL is always
satisfied in apparent horizon, which do
not depend on BD theory of gravity.\\

{\bf Acknowledgement:}\\

The authors are thankful to IUCAA, Pune, India for warm
hospitality where the work was done.\\

{\bf References:}\\
\\
$[1]$ S. J. Perlmutter et al, {\it Bull. Am. Astron. Soc.} {\bf
29} 1351 (1997); S. J. Perlmutter et al, \textit{Astrophys. J.} \textbf{517} 565 (1999).\\
$[2]$ A. G. Riess et al, {\it Astron. J.} {\bf 116} 1009 (1998);
P. Garnavich et al, {\it Astrophys. J.} {\bf 493} L53 (1998); B.
P. Schmidt et al, {\it Astrophys. J.} {\bf 507} 46 (1998); N. A.
Bachall, J. P. Ostriker, S. Perlmutter and P. J. Steinhardt,
\textit{Science} \textbf{284} 1481 (1999).\\
$[3]$ R. R. Caldwell, R. Dave and P. J. Steinhardt, {\it Phys. Rev. Lett.} {\bf 80} 1582 (1998).\\
$[4]$ A. S. Al-Rawaf and M. O. Taha, {\it Gen. Rel. Grav.} {\bf
28} 935 (1996).\\
$[5]$ T. Padmanabhan, {\it Phys. Rept.} {\bf 380} 235 (2003).\\
$[6]$ N. Banerjee and D. Pavon, {\it Phys. Rev. D} {\bf 63} 043504 (2001).\\
$[7]$ B. K. Sahoo and L. P. Singh, {\it Modern Phys. Lett. A} {\bf 18} 2725- 2734 (2003).\\
$[8]$ K. Nordtvedt, Jr., {\it Astrophys. J} {\bf 161} 1059 (1970);
P. G. Bergmann, {\it Int. J. Phys.} {\bf 1} 25 (1968);
R. V. Wagoner, {\it Phys. Rev. D} {\bf 1} 3209 (1970).\\
$[9]$ O. Bertolami and P. J. Martins, {\it Phys. Rev. D} {\bf 61} 064007 (2000).\\
$[10]$ R. Ritis, A. A. Marino, C. Rubano and P. Scudellaro, {\it
Phys. Rev. D} {\bf 62} 043506 (2000); B. Boisseau, G.
Esposito-Farese, D. Polarski and A. A. Starobinsky, {\it Phys.
Rev. Lett.} {\bf 85} 2236 (2000).\\
$[11]$ S. Sen and A. A. Sen, {\it Phys. Rev. D} {\bf 63} 124006
(2001); S. Sen and T. R. Seshadri, {\it Int. J. Mod. Phys. D} {\bf
12} 445 (2003); W. Chakraborty and U. Debnath, {\it Int. J. Theor.
Phys.} {\bf 48} 232 (2009); V. Faraoni, {\it Phys. Rev. D} {\bf
62} 023504 (2000); T. D. Saini, S. Raychauchaudhury, V. Sahni and
A. A. Starobinsky, {\it Phys. Rev. Lett.} {\bf 85} 1162 (2000).\\
$[12]$ T. Jacobson, {\it Phys. Rev. Lett.} {\bf 75} 1260 (1995).\\
$[13]$ J. D. Bekenstein, {\it Phys. Rev. D} {\bf 7} 2333 (1973);
S. W. Hawking, {\it Commun. Math. Phys.} {\bf 43} 199 (1975); J.
M. Bardeen, B. Carter and S. W. Hawking, {\it Commun. Math. Phys.}
{\bf 31} 161 (1973).\\
$[14]$ T. Padmanabhan, {\it Class. Quantum Grav.} {\bf 19} 5387
(2002).\\
$[15]$ A. V. Frolov and L. Kofman, {\it JCAP} {\bf 0305} 009
(2003).\\
$[16]$ R. G. Cai and S. P. Kim, {\it JHEP} {\bf 02} 050 (2005).\\
$[17]$ G. W. Gibbons and S. W. Hawking, {\it Phys. Rev. D} {\bf 15} 2738 (1977).\\
$[18]$ B. Wang, Y. G. Gong and E. Abdalla, {\it Phys. Rev. D} {\bf 74} 083520 (2006).\\
$[19]$ Y. Gong, B. Wang and A. Wang, {\it JCAP} {\bf 01} 024 (2007).\\
$[20]$ S. -F. Wu, B. Wang and G. -H Yang, {\it Nucl. Phys. B} {\bf
799} 330 (2008).\\
$[21]$ E. Babichev, V. Dokuchaev and Yu. Eroshenko, {\it Phys.
Rev. Lett.} {\bf 93} 021102 (2004); M. R. Setare and S. Shafei,
{\it JCAP} {\bf 09} 011 (2006); M. R. Setare, {\it Phys. Lett. B}
{\bf 641} 130 (2006); P. C. W. Davies, {\it Class. Quant. Grav.}
{\bf 4} L225 (1987); H. M. Sadjadi, {\it Phys. Rev. D} {\bf 73}
063525 (2006).\\
$[22]$ G. Izquierdo, D. Pavon, {\it Phys. Lett. B} {\bf 639} 1
(2006); H. M. Sadjadi, {\it Phys. Lett. B} {\bf 645} 108 (2007).\\
$[23]$ S. -F. Wu, B. Wang, G. -H. Yang and P. -M. Zhang,
{\it Class. Quant. Grav.} {\bf 25} 235018 (2008).\\
$[24]$ A. Sheykhi and B. Wang, {\it Phys. Lett. B} {\bf 678} 434 (2009).\\
$[25]$ R. -G. Cai and N. Ohta, {\it Phys. Rev. D} {\bf 81} 084061
(2010); N. Mazumder and S. Chakraborty, arXiv: 1003.1606[gr-qc];
M. Jamil, A. Sheykhi and M. U. Farooq, arXiv:1003.2093[hep-th]; A.
Wang and Y. Wu, {\it JCAP} {\bf 0907} 012 (2009); Q. -J. Cao, Y.
-X. Chen and K. -N. Shao, arXiv:1001:2597[hep-th].\\
$[26]$ R. -G. Cai, L. -M. Cao, Y. -P. Hu and S. P. Kim, {\it Phys.
Rev. D} {\bf 78} 124012 (2008).\\
$[27]$ D. Mateos, R. C. Myers and R. M. Thomson, {\it JHEP} {\bf
05} 067 (2007); R. -G. Cai, {\it Prog. Theor. Phys.} {\bf 172} 100
(2008);
R. G. Cai and L. -M. Cao, {\it Nucl. Phys. B} {\bf 785} 135 (2007).\\
$[28]$ M. Akbar and R. -G. Cai, {\it Phys. Lett. B} {\bf 635} 7 (2006).\\
$[29]$ H. M. Sadjadi, {\it Phys. Rev. D} {\bf 76} 104024 (2007);
Q. -J. Cao, Y. -X. Chen and K. -N. Shao, arXiv:1001.2597.\\
$[30]$ C. Brans and R. H. Dicke, {\it Phys. Rev.} {\bf 124} 925
(1961).\\
$[31]$ M. G. Green, J. H. Schwarz and E. Witten, {\it Superstring
theory} (Cambridge Univ. Press, Cambridge, 1987).\\
$[32]$ N. Mazumder and S. Chakraborty, {\it Class. Quantum Grav.}
{\bf 26} 195016 (2009).\\
$[33]$ R. S. Bousso {\it Phys. Rev. D} {\bf 71} 064024 (2005).\\

\end{document}